\documentclass[runningheads]{llncs}
\usepackage{graphicx}
\usepackage[para,online,flushleft]{threeparttable}
\usepackage{booktabs}
\usepackage{multirow}
\usepackage{lipsum}
 
\newcommand\blfootnote[1]{%
  \begingroup
  \renewcommand\thefootnote{}\footnote{#1}%
  \addtocounter{footnote}{-1}%
  \endgroup
}
\begin{document}
\title{PASH at TREC 2021 Deep Learning Track: Generative Enhanced Model for Multi-stage Ranking}
%
%
\author{Yixuan Qiao\inst{1,\footnotesize{*}} \and
Shanshan Zhao\inst{2} \and
Jun Wang\inst{1} \and
Hao Chen\inst{1} \and
Xin Tang\inst{3} \and
Tuozhen Liu\inst{4} \and 
Xianbin Ye\inst{1} \and 
Rui Fang\inst{3} \and
Wenfeng Xie\inst{3} \and
Guotong Xie\inst{1,5,6,\footnotesize{*}}
}
%
\authorrunning{Y.X. Qiao et al.}
%
\institute{Ping An Health Technology, Beijing, China 
\and
Peking University, Beijing, China
\and
Ping An Property \& Casualty Insurance Company, Shenzhen, China
\and
Central University of Finance and Economics, Beijing, China
\and
Ping An Health Cloud Company Limited., Shenzhen, China \and
Ping An International Smart City Technology Co., Ltd., Shenzhen, China
}
\maketitle              
\blfootnote{$^*$Corresponding Author. Email: \protect\url{\{qiaoyixuan528, xieguotong\}@pingan.com.cn}.}
\begin{abstract}
This paper describes the PASH participation in TREC 2021 Deep Learning Track. In the recall stage, we adopt a scheme combining sparse and dense retrieval method. In the multi-stage ranking phase, point-wise and pair-wise ranking strategies are used one after another based on model continual pre-trained on general knowledge and document-level data. Compared to TREC 2020 Deep Learning Track, we have additionally introduced the generative model T5 to further enhance the performance. 

\end{abstract}

\section{Introduction}

TREC 2021 is the thirtieth edition of the Text REtrieval Conference (TREC). The main goal of TREC is to create the evaluation infrastructure required for large-scale testing of information retrieval (IR) technology. This includes research on best methods for evaluation as well as development of the evaluation materials themselves. “Retrieval technology” is broadly interpreted to include a variety of techniques that enable and/or facilitate access to information that is not specifically structured for machine use. Specifically, the Deep Learning track focuses on IR tasks where a large training set is available, allowing us to compare a variety of retrieval approaches including deep neural networks and strong non-neural approaches, to see what works best in a large-data regime.

The Deep Learning track had two tasks, Document Ranking and Passage Ranking. For each task, participants could either do their own retrieval from the full collection or re-rank an initial retrieved set provided by the track organizers. Both tasks used the same set of 477 test questions.

\section{Methodology}
In this section, we briefly describe the components in the multi-stage ranking pipeline. Details of some methods can be found in TREC 2020 notebook paper. The passage ranking task contains 285,328 queries on a collection of 138,364,198 passages, totally 296,419 query passage pairs annotated as positive for relevance. Few queries are matched with multiple relevant passages. The document ranking task contains 331,748 queries on a collection of 11,959,635 documents, totally 341,836 query document pairs annotated as positive for relevance. Few queries are matched with multiple relevant documents. NDCG (Normalized Discounted Cumulative Gain) is the main official evaluation index. 

\subsection{Multi-way Matching}
\subsubsection{Sparse retrieval} We use the docTTTTTquery\cite{nogueira2019document} (also known as docT5query or doc2query-T5) to generate queries for which the passage might be relevant. For passage, We sample 40 queries per passage with T5-base. For document, we first segment each document into passages by applying a sliding window of ten sentences with a stride of five. Each passage is then prepended with the url and title of the document. Finally, we generate ten queries per passage. Each document prepended with the url and title info was appended with all its predicted queries are indexed by BM25\cite{zeng2019bm25} as before.

\subsubsection{Dense retrieval} Dense retrieval methods have shown great promise over sparse retreval methods in a range of NLP (Natural Language Processing) problems. We choose ColBERT\cite{khattab2020colbert} as our Dense Retriever due to the effectiveness/efficiency trade-offs.

\subsection{Multi-stage Ranking}
We inherit the model from TREC 2020 Deep Learning Track including BERT-Large\cite{devlin2019bert},  ALBERT-XXLarge\cite{lan2019albert}, ELECTRA-Large\cite{clark2020electra} and XLNet-Large\cite{yang2019xlnet} model. In particular, BERT-Large uses the same pre-training and fine-tuning strategies as last year but using this year’s data set. Compared with last year, we have made the following further improvements:

\begin{itemize}
    \item We use pair-wise loss as the goal of the second stage of ranking. So we can use more data instead of TREC labeled 4-class data. 
    \item Generative model T5\cite{raffel2020exploring} is introduced to further enhance the performance
    \begin{itemize}
        \item[a)] Implementation based on Megatron-LM\cite{narayanan2021efficient} repository.
        \item[b)] Same training strategy as \cite{pradeep2021expandomonoduo}.
        \item[c)] Training the T5-3B model takes approximately 12 days on 32 V100 GPUs consuming about 12.8M training examples.
        \item[d)] The T5-11B model takes about 30 days under the same situation. An amazing fact is that the zero-shot NDCG@5 of the T5-11B is already 0.3651.
    \end{itemize}
\end{itemize} 

\subsection{Ensemble} Benefited from multi-way matching, we just sequentially train different re-rankers with different random seeds for ensemble learning.

\section{Results}
We submitted six official runs in each passage and document ranking sub-task. Table \ref{p_r} and \ref{p_f} present the results of our runs in passage ranking task. Table \ref{d_r} and \ref{d_f} present the results of our runs in document ranking task. We use $\_$r${\ast}$ to denote the runs for re-ranking, $\_$f${\ast}$ for full ranking.


\begin{table*}[hp]
\centering
    \caption{Results on the TREC 2021 passage re-rank task.}
    \label{p_r}
      \begin{threeparttable}
    \begin{tabular}{lcccc}
    \toprule
    RUN & MAP@100 & NDCG@5 & NDCG@10 & R@100\\ 
   \hline
    pash\_r1 & 0.2362 & 0.7190 & 0.6951 & 0.3261 \\
    \hline
    pash\_r2 & 0.2389 & 0.7390 & 0.7076 & 0.3261 \\
    \hline
    pash\_r3 & 0.2385 & 0.7390 & 0.7072 & 0.3261 \\
    \bottomrule
    \end{tabular}
    \footnotesize
    \end{threeparttable}
\end{table*}

\begin{table*}[hp]
\centering
    \caption{Results on the TREC 2021 passage full rank task.}
    \label{p_f}
      \begin{threeparttable}
    \begin{tabular}{lcccc}
    \toprule
    RUN & MAP@100 & NDCG@5 & NDCG@10 & R@100\\ 
   \hline
    pash\_f1 & 0.3193 & 0.7596 & 0.7494 & 0.4868 \\
    \hline
    pash\_f2 & 0.3318 & 0.7596 & 0.7494 & 0.5300 \\
    \hline
    pash\_f3 & 0.3378 & 0.7596 & 0.7494 & 0.5500 \\
    \bottomrule
    \end{tabular}
    \footnotesize
    \end{threeparttable}
\end{table*}

\begin{table*}[hp]
\centering
    \caption{Results on the TREC 2021 document re-rank task.}
    \label{d_r}
      \begin{threeparttable}
    \begin{tabular}{lcccc}
    \toprule
    RUN & MAP@100 & NDCG@5 & NDCG@10 & R@100\\ 
   \hline
    pash\_doc\_r1 & 0.2665 & 0.7452 & 0.7150 & 0.3195 \\
    \hline
    pash\_doc\_r2 & 0.2640 & 0.7308 & 0.7076 & 0.3195 \\
    \hline
    pash\_doc\_r3 & 0.2672 & 0.7383 & 0.7164 & 0.3195 \\
    \bottomrule
    \end{tabular}
    \footnotesize
    \end{threeparttable}
\end{table*}

\begin{table*}[ht!]
\centering
    \caption{Results on the TREC 2021 document full rank task.}
    \label{d_f}
      \begin{threeparttable}
    \begin{tabular}{lcccc}
    \toprule
    RUN & MAP@100 & NDCG@5 & NDCG@10 & R@100\\ 
   \hline
    pash\_doc\_f1 & 0.3111 & 0.6745 & 0.5832 & 0.3683 \\
    \hline
    pash\_doc\_f4 & 0.3498 & 0.7516 & 0.7404 & 0.4412 \\
    \hline
    pash\_doc\_f5 & 0.3521 & 0.7508 & 0.7368 & 0.4419 \\
    \bottomrule
    \end{tabular}
    \footnotesize
    \end{threeparttable}
\end{table*}



%
%

\section{Conclusion}
The experiments demonstrate that the expressiveness of ColBERT is well integrated with BM25+docTTTTTquery. Compared with the transformer-based model, the generative model T5 has a higher NDCG@5 and a lower NDCG@10, which means that relevant passages are already in top5. This phenomenon largely hints at the thirst for multi-stage ranking in generative models, which can be expected to show powerful effects with the pair-wise or list-wise training. We believe that the interweaving sequence of different methods would be a focus priority in the future work. 

%
%
%
%






\bibliographystyle{naturemag}
\bibliography{ref}

\end{document}